\documentclass[12pt]{article}
\usepackage{graphics} 
\usepackage{cite}
\usepackage{epsfig}
\newcommand  {\etal}     {{\it et al.}}

\newcommand  {\Bioch}    {{\it Biochemistry\ }}

\newcommand  {\Biopol}   {{\it Biopolymers\ }}

\newcommand  {\BJ}       {{\it Biophys.~J.\ }}

\newcommand  {\COSB}     {{\it Curr.\ Opin.\ Struct.\ Biol.\ }}

\newcommand  {\EL}       {{\it Europhys.\ Lett.\ }}

\newcommand  {\JBP}      {{\it J.\ Biol.\ Phys.\ \ }}

\newcommand  {\JCP}      {{\it J.\ Chem.\ Phys.\ }}
\newcommand  {\JMB}      {{\it J.\ Mol.\ Biol.\ }}

\newcommand  {\Nat}      {{\it Nature\ }}
\newcommand  {\NSB}      {{\it Nat.\ Struct.\ Biol.\ }}

\newcommand  {\Pro}      {{\it Proteins\ }}

\newcommand  {\ProSci}   {{\it Protein\ Sci.\ }}

\newcommand  {\PNAS}     {{\it Proc.\ Natl.\ Acad.\ Sci.\ USA\ }}

\newcommand  {\Sci}      {{\it Science\ }}

\newcommand{\beq}{\begin{equation}}
\newcommand{\eeq}{\end{equation}}
\newcommand{\beqa}{\begin{eqnarray}}
\newcommand{\eeqa}{\end{eqnarray}}
\newcommand{\bea}{\begin{eqnarray}}
\newcommand{\eea}{\end{eqnarray}}

\newcommand   {\Ca}      {C${}_{\alpha}$}
\newcommand   {\Cb}      {C${}_{\beta}$}
\newcommand   {\Cd}      {C${}_{\delta}$}
\newcommand   {\Cp}      {C${}^{\prime}$}

\newcommand   {\Eloc}    {E_{\mbox{{\scriptsize loc}}}}
\newcommand   {\Eev}     {E_{\mbox{{\scriptsize ev}}}}
\newcommand   {\Ehb}     {E_{\mbox{{\scriptsize hb}}}}

\newcommand   {\Ehp}     {E_{\mbox{{\scriptsize hp}}}}
\newcommand   {\Tm}      {T_{\mbox{{\scriptsize m}}}}
\newcommand   {\Nh}      {N_{\mbox{{\scriptsize h}}}}

\newcommand   {\Rg}      {R_{\mbox{{\scriptsize g}}}}
\newcommand   {\Zspa}    {Z${}_{{\rm SPA-1}}$} 

\input{psfig}

\begin{document}

\begin{flushright}
LU TP 03-14\\
June 25, 2003
\end{flushright}

\vspace{0.4in}

\begin{center}

{\LARGE \bf Sequence-Based Study of Two Related} 

{\LARGE \bf Proteins with Different Folding Behaviors} 

\vspace{.6in}

\large
Giorgio Favrin, Anders Irb\"ack and Stefan 
Wallin\footnote{E-mail: favrin,\,anders,\,stefan@thep.lu.se}\\   
\vspace{0.10in}
Complex Systems Division, Department of Theoretical Physics\\ 
Lund University,  S\"olvegatan 14A,  SE-223 62 Lund, Sweden \\
{\tt http://www.thep.lu.se/complex/}\\

\vspace{0.3in}	

\end{center}
\vspace{0.2in}
\normalsize
Abstract:\\
\Zspa\ is an engineered protein that binds to its parent, the
three-helix-bundle Z domain of staphylococcal protein A.
Uncomplexed \Zspa\ shows a reduced helix content and a melting
behavior that is less cooperative, compared
with the wild-type Z domain. Here we show that the difference in
folding behavior between these two sequences can be partly
understood in terms of an off-lattice model with
5--6 atoms per amino acid and a minimalistic potential,
in which folding is driven by backbone hydrogen bonding and
effective hydrophobic attraction.

\vspace{24pt}

Keywords: protein folding, folding thermodynamics, three-helix bundle,
unstructured protein, Monte Carlo simulation. 

\newpage 

\section{Introduction}

It is becoming increasingly clear that unstructured proteins play an
important biological role~\cite{Wright:99,Dyson:02}. In many cases,
such proteins adopt a specific structure upon binding to their
biological targets. Recently, it was demonstrated that the {\it in vitro}
evolved \Zspa\ protein~\cite{Eklund:02} exhibits coupled folding and
binding~\cite{Wahlberg:03}.
 
\Zspa\ is derived from the Z domain of staphylococcal protein A,
a 58-amino acid, well characterized~\cite{Tashiro:97} three-helix-bundle
protein. \Zspa\ was engineered~\cite{Eklund:02}
by randomizing 13 amino acid positions and
selecting for binding to the Z domain itself. Subsequently,
the structure of the Z:\Zspa\ complex was determined both in
solution~\cite{Wahlberg:03} and by crystallography~\cite{Hogbom:03}.
In the complex, both \Zspa\ and the Z domain adopt structures similar to
the solution structure of the Z domain. However, in solution,
\Zspa\ does not behave as the Z domain; Wahlberg~\etal~\cite{Wahlberg:03}
found that uncomplexed \Zspa\ lacks a well-defined structure,
and that its melting behavior is less cooperative than that of
the Z domain.
 
The Z domain is a close analog of the B domain of protein A, a chain
that is known to show two-state folding without any meta-stable intermediate
state~\cite{Bai:97,Myers:01}. The folding behavior of the B domain has
been studied theoretically by many different groups,
including ourselves, using both
all-atom~\cite{Boczko:95,Guo:97,Kussell:02,Linhananta:02} and
reduced~\cite{Kolinski:98,Zhou:99,Shea:99,Berriz:01,Favrin:02} models.
In many cases, it was possible to fold this chain, but
to achieve that most models rely on the so-called G\=o
prescription~\cite{Go:81}.
Our model~\cite{Favrin:02} is, by contrast,
sequence-based.
This makes it possible for us to study
both \Zspa\ and the wild-type Z domain and compare
their behaviors, using one and the same model.
 
The purpose of this note is twofold. First, we check whether our model
can explain the difference in melting behavior
between \Zspa\ and the wild-type sequence. Second,
using this model, we study the structure of \Zspa.

\section{Materials and Methods}\label{sec:mod}

\subsection{Model}
 
The model we study~\cite{Favrin:02} is an extension of a model with three
amino acids~\cite{Irback:00,Irback:01,Favrin:03} to a five-letter
alphabet. The five amino acid types are hydrophobic (Hyd),
polar (Pol), Ala, Pro and Gly.
Hyd, Pol and Ala share the same geometric representation but differ
in hydrophobicity. Pro and Gly have their own geometric representations.

The Hyd, Pol and Ala representation contains six atoms. The three backbone
atoms N, \Ca\ and \Cp\ and the H and O atoms of the peptide unit are all
included. The H and O atoms are used to define hydrogen bonds. The sixth atom
is a large \Cb\ that represents the side chain. Gly lacks the \Cb\ atom
but is otherwise the same. The representation of Pro differs from
that of Hyd, Pol and Ala in that the H atom is replaced by a side-chain
atom, \Cd, and that the Ramachandran angle $\phi$ is held fixed at
$-65^\circ$.

The degrees of freedom of our model are the Ramachandran torsion angles
$\phi$ and $\psi$, with the exception
that $\phi$ is held fixed for Pro. All bond lengths, bond angles and
peptide torsion angles ($180^\circ$) are held fixed.

The interaction potential
\beq
E=\Eloc+\Eev+\Ehb+\Ehp
\label{e}\eeq
is composed of four terms. The first term is a local $\phi,\psi$ potential.
The other three terms represent excluded volume, backbone hydrogen bonds
and effective hydrophobic attraction, respectively (no explicit water).
For simplicity, the hydrophobicity potential is taken to be pairwise
additive. Only Hyd-Hyd and Hyd-Ala \Cb\ pairs experience this type of
interaction. In particular, this means that Ala is intermediate in
hydrophobicity between Hyd and Pol. The amino acids in the Hyd class
are Val, Leu, Ile, Phe, Trp and Met, whereas those in the
Pol class are Arg, Asn, Asp, Cys, Gln, Glu, His, Lys, Ser, Thr and  Tyr.
A complete description of the model, including numerical values of
all the parameters, can be found in our earlier study~\cite{Favrin:02}.
 
In this earlier study, the model was applied to the 10--55-amino acid
fragment of the B domain of protein A. Despite the simplicity of the
potential, this sequence was found to have the following
properties~\cite{Favrin:02} in the model:
\begin{itemize}
\item It does make a three-helix bundle with the native three-helix-bundle
topology,\footnote{There are two possible three-helix-bundle
topologies; if we let the first two helices form a U, then the third helix
can be either in front of or behind this U.}
although the suppression of the wrong topology is not very strong. All
helices are right-handed, as they should.
\item Energy minimization restricted to the thermodynamically favored
(native) topology gives a structure with a root-mean-square deviation (RMSD)
of 1.8\,\AA\ from the NMR structure~\cite{Gouda:92} (calculated over
all backbone atoms).
\item The collapse transition is much more cooperative for this sequence
than for random sequences with the same composition. Moreover, chain
collapse and helix formation occur at approximately the same temperature.
\end{itemize}
The relative order of chain collapse and helix formation depends strongly
on the relative strength of the hydrogen bonds and the hydrophobic attraction,
so the last conclusion may seem somewhat arbitrary. However, the chain
does not fold to a helical bundle if the hydrogen bonds are too strong,
and it does not fold in a cooperative manner if the hydrogen bonds are too
weak~\cite{Irback:01}. As a result, with our ansatz for the potential, there
is not much freedom left in the choice of these parameters, if the chain is to
fold to a compact helical bundle in a cooperative manner.
 
In the present study, we apply the same model, with unchanged parameters,
to \Zspa\ and the Z domain of protein A. Following previous calculations for
the B
domain~\cite{Boczko:95,Guo:97,Linhananta:02,Kolinski:98,Zhou:99,Shea:99,Berriz:01,Favrin:02}, we consider the 9--54-amino acid fragments of these two
sequences (corresponding to the 10--55-amino acid fragment of the
B domain).
It should be mentioned that we also performed calculations for
the 4--54-amino acid fragments of \Zspa\ and the Z domain,
with similar results.
 
The amino acid sequences of \Zspa\ and the Z domain differ at 13 positions,
all of which are located in the section 9--35. Table~I shows this part of the
sequences.

\begin{table}[t]
\begin{center}
\begin{tabular}{llllllllll}
Z          & QQN & AFY & EIL & HLP & NLN & EEQ & RNA & FIQ & SLK\\
\Zspa      & LSV & AGR & EIV & TLP & NLN & DPQ & KKA & FIF & SLW        
\end{tabular}
\caption{Amino acids 9 to 35 for \Zspa\ and the wild-type 
Z domain.}
\label{tab:1}
\end{center}
\end{table}
    
\subsection{Numerical Methods} 

To simulate the thermodynamic behavior of this model, we use
simulated tempering~\cite{Lyubartsev:92,Marinari:92}, in which the
temperature is a dynamic variable.
Details on our implementation of this method can be
found elsewhere~\cite{Irback:95}.
For a review of simulated tempering and other
generalized-ensemble techniques,
see~\cite{Hansmann:99}.
 
In conformation space we use two different elementary moves:
first, the pivot move in which a single torsion angle
is turned; and second, a semi-local method~\cite{Favrin:01}
that works with seven or eight adjacent torsion angles, which are turned
in a coordinated manner. The non-local pivot move is included in our
calculations in order to accelerate the evolution
of the system at high temperature, whereas the semi-local method
improves the performance at low temperature.
 
Our simulations
are started from random configurations.
All statistical errors quoted are 1$\sigma$
errors obtained by analyzing data from eight independent runs.
 
The temperatures
studied range from $0.87\,\Tm$ to $1.43\,\Tm$, $\Tm$ being the
melting temperature for the wild-type Z domain. The experimental
value of this temperature is $\Tm=75^\circ$C~\cite{Wahlberg:03}.
Hence, the lowest and highest temperatures studied correspond to
$31^\circ$C and $225^\circ$C, respectively. In the dimensionless
energy unit used in our earlier study~\cite{Favrin:02},
$\Tm$ is given by $k\Tm=0.630\pm0.001$, $k$ being Boltzmann's constant.
In the model we define $\Tm$ as the maximum of the specific heat.

\section{Results and Discussion}\label{sec:res}

\begin{figure}
\vspace{0mm}
\begin{center}
\mbox{
\epsfig{figure=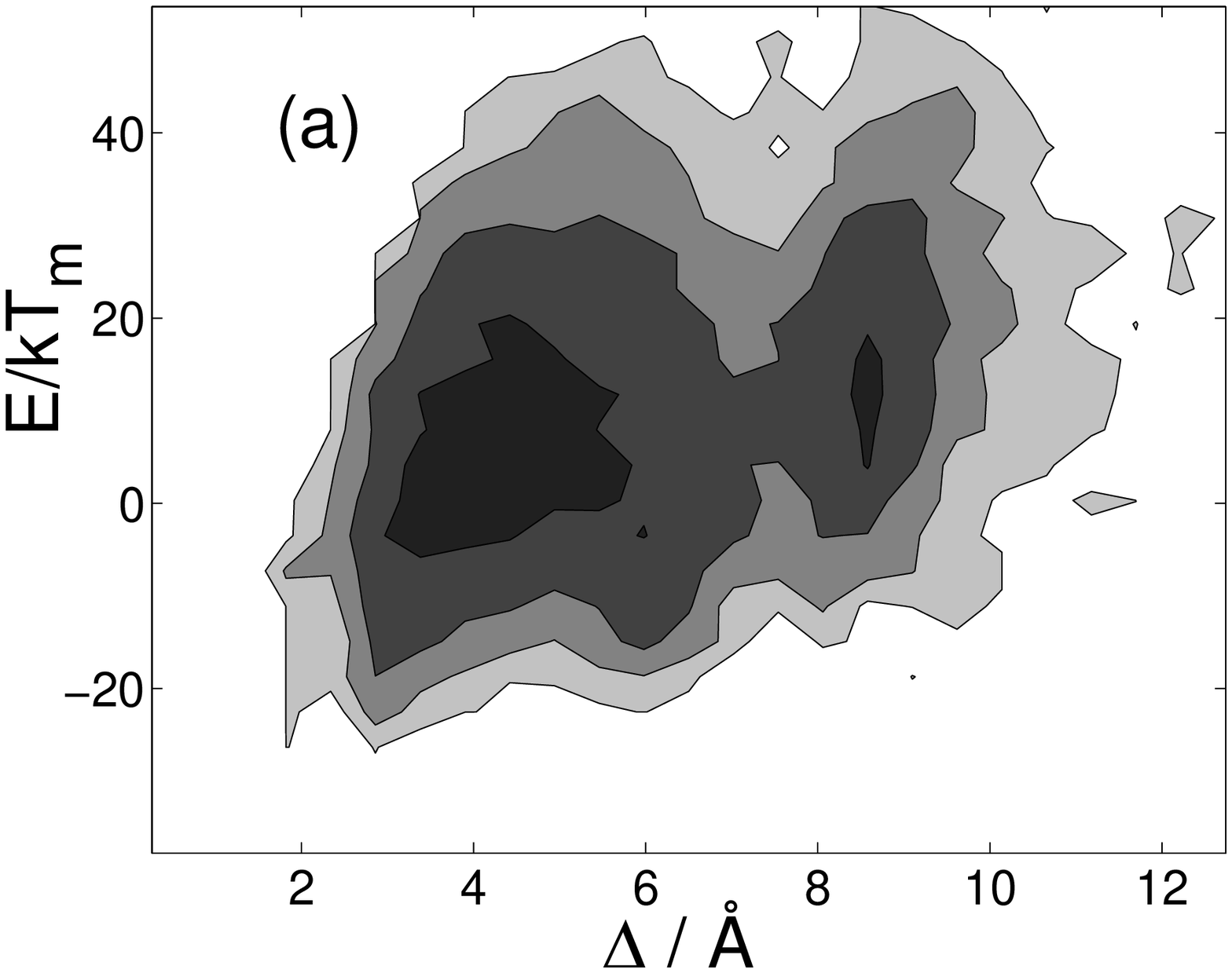,height=5.5cm}
\hspace{5mm}
\epsfig{figure=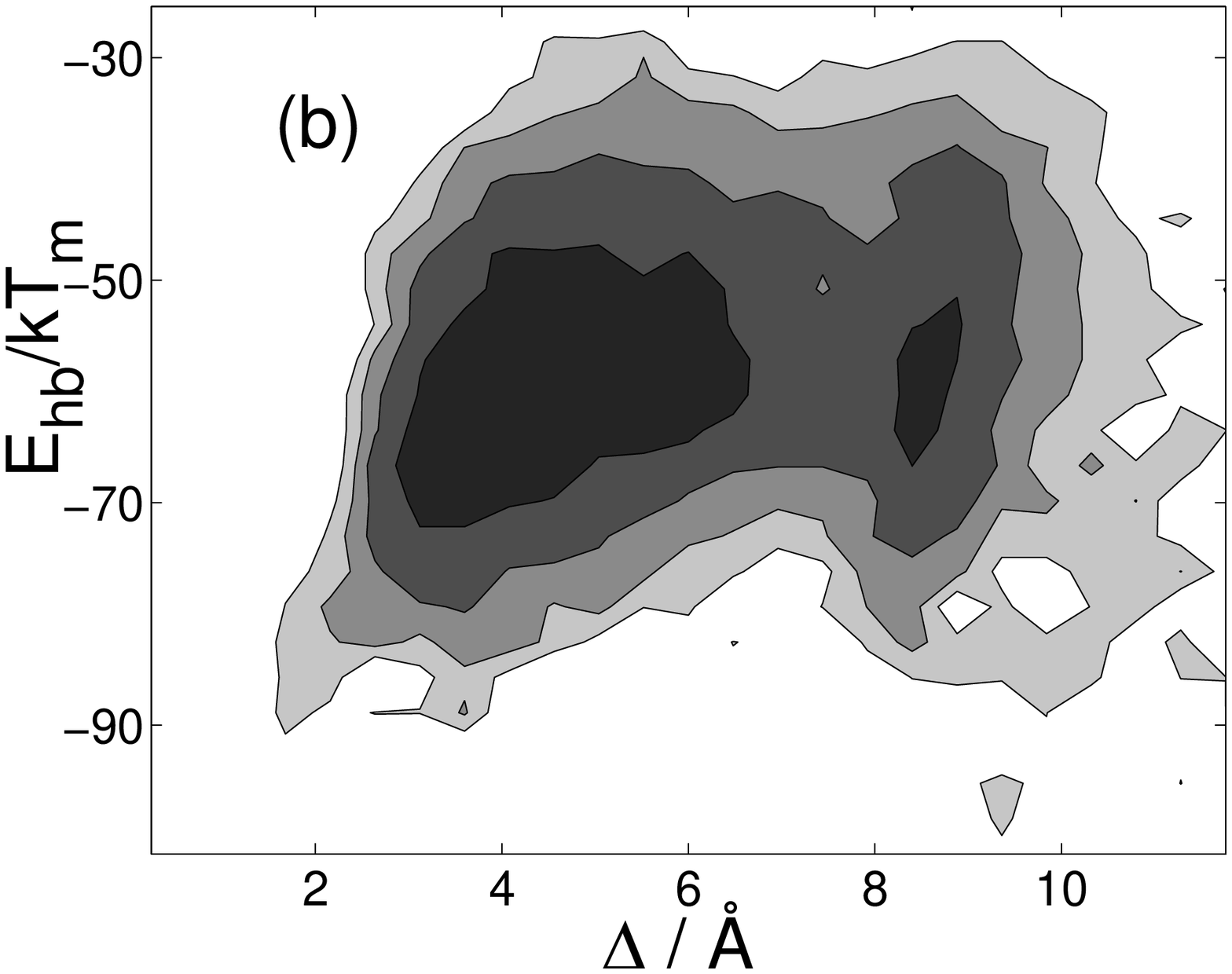,height=5.5cm}
}
\end{center}
\vspace{-12pt}
\caption{
Level diagrams showing the free energies (a) $F(\Delta,E)$  and
(b) $F(\Delta,\Ehb)$ for the 9--54-amino acid fragment
of the Z domain at $T=0.87\,\Tm$. $E$ is the total energy [see
equation (\ref{e})], $\Ehb$ is the hydrogen-bond energy, and
$\Delta$ denotes RMSD from the NMR structure. The separation between
adjacent contour lines is 1\,$kT$. The darkest regions correspond
to $4<F/kT<5$ and the white regions to $F/kT>8$.}
\label{fig:1}
\end{figure}

\begin{figure}[t]
\vspace{0mm}
\begin{center}
\epsfig{figure=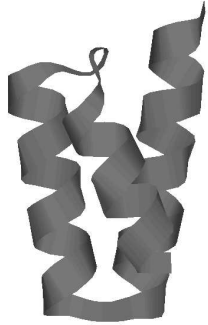,height=4.5cm}
\end{center}
\vspace{-12pt}
\caption{Schematic illustration of the structure obtained
by energy minimization restricted to the thermodynamically favored
topology (see text) for the 9--54-amino acid fragment of the Z domain.}
\label{fig:2}
\end{figure}

Using the model and methods described in the previous section, we study
the 9--54-amino acid fragments of \Zspa\ and the wild-type Z domain.
The latter sequence differs only by a one-point mutation
from the previously studied 10--55-amino acid fragment of the B
domain. Our results for the Z domain are similar to those
for the B domain~\cite{Favrin:02} summarized in the previous section.
Figure~\ref{fig:1} shows the free energies $F(\Delta,E)$ and $F(\Delta,\Ehb)$
for the Z domain at $T=0.87\,\Tm$, where $\Delta$ denotes
RMSD from the NMR structure~\cite{Tashiro:97} (PDB code 2SPZ, model 1).
Two major minima can be seen, with similar hydrogen-bond energies.
These minima correspond to the two possible three-helix-bundle topologies.
Both topologies are significantly populated, but the average total energy
is slightly lower for the native topology, and this topology is the
thermodynamically favored one.
We also performed an energy minimization for the native topology,
by applying simulated annealing combined with
a conjugate gradient method to a large number of low-temperature
conformations. The minimum-energy structure obtained this way is
schematically illustrated in Fig.~\ref{fig:2}. It has an RMSD 
of $\Delta=1.7$\,\AA\
from the NMR structure. The corresponding result for the B domain was,
as mentioned earlier, $\Delta=1.8$\,\AA.

\begin{figure}
\vspace{0mm}
\begin{center}
\mbox{
\epsfig{figure=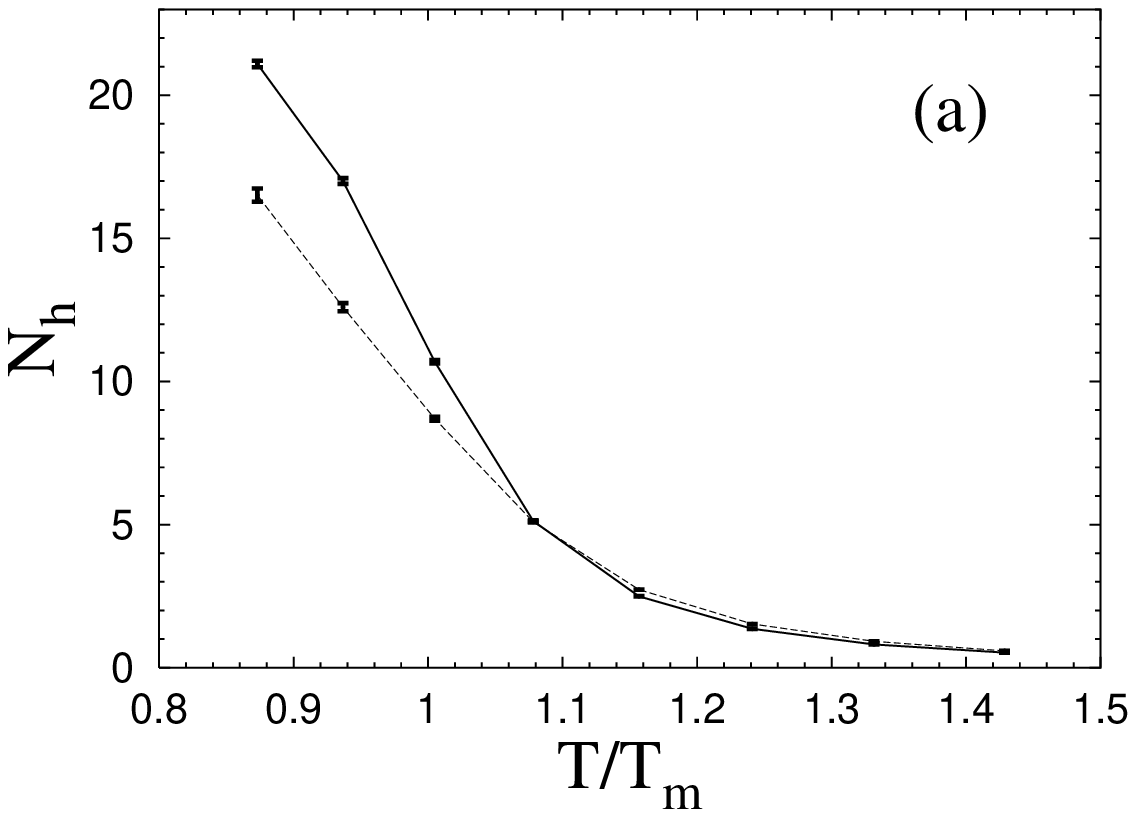,height=5.5cm}
\hspace{5mm}
\epsfig{figure=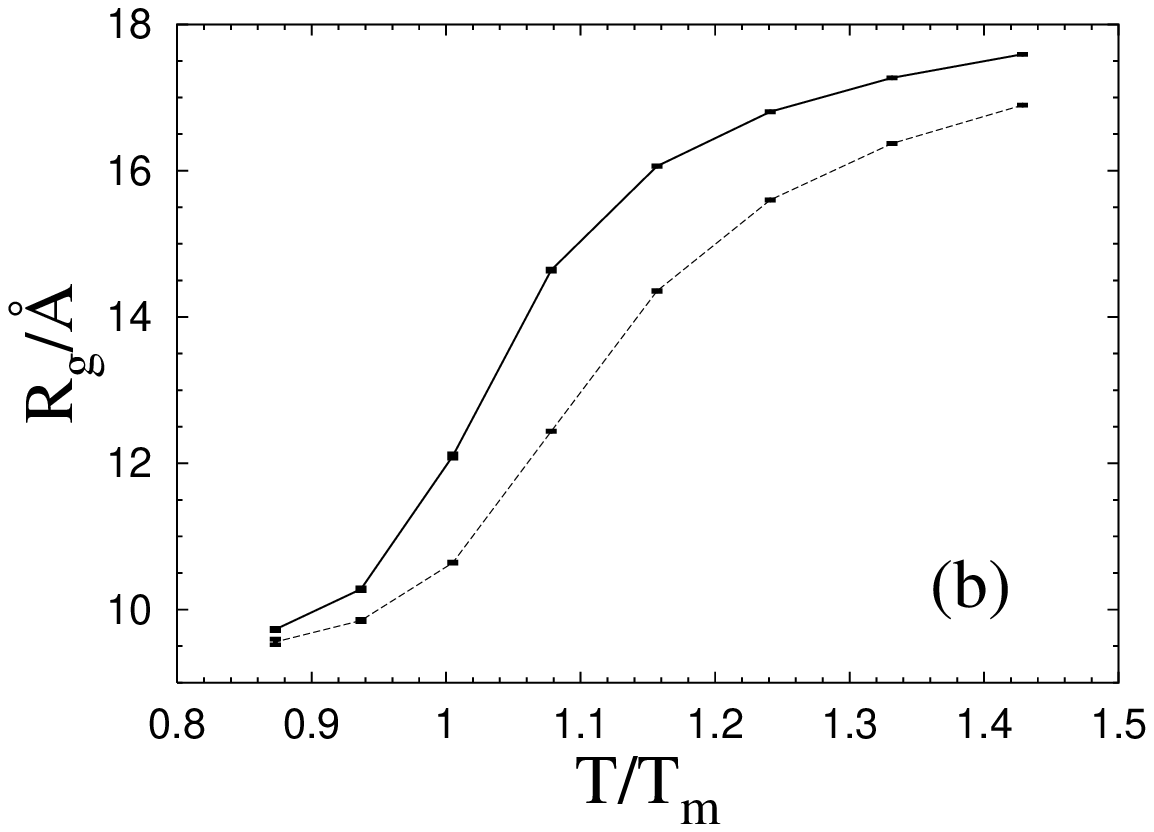,height=5.5cm}
}
\end{center}
\vspace{-12pt}
\caption{
Helix formation and chain collapse for the 46-amino acid fragments
of \Zspa\ (dashed line) and the Z domain (full line). (a) The number
of helical amino acids, $\Nh$,  against temperature.
(b) The radius of gyration (calculated over all backbone atoms), $\Rg$,
against temperature. $\Tm$ denotes the melting temperature for the
Z domain. The NMR structure for the Z domain has $\Nh=29$ and
$\Rg=9.0$\,\AA.}
\label{fig:3}
\end{figure}

Let us now compare the behavior of the Z domain with that of the
engineered \Zspa\ sequence. By CD, Wahlberg~\etal~\cite{Wahlberg:03}
found \Zspa\ to be less helical than the wild-type Z domain, the mean residue
ellipticity for \Zspa\ being 60\% of that for the wild-type sequence.
Furthermore, they found that the helix formation sets in at a lower
temperature and is less cooperative for the engineered sequence.
Figure~\ref{fig:3}a shows the helix content against temperature in our model,
for both
sequences.\footnote{We define helix content in the following way.
Each amino acid, except the two at the ends, is labeled h
if $-90^\circ<\phi<-30^\circ$ and $-77^\circ<\psi<-17^\circ$,
and c otherwise. The two amino acids at the ends are labeled c.
An amino acid is said to be helical if both the amino acid itself
and its nearest neighbors are labeled h. The total number of
helical amino acids is denoted by $\Nh$. The maximum value of
$\Nh$ is $N-4$ for a chain with $N$ amino acids.}
In agreement with the experimental results, we find that \Zspa\
has a lower helix content, and that the helix formation is
shifted toward lower temperature for this sequence.
Figure~\ref{fig:3}b shows the temperature dependence of the radius of
gyration. We find that \Zspa\ is more compact than the
Z domain. A comparison with Fig.~\ref{fig:3}a shows that, in our model,
chain collapse occurs before helix formation for \Zspa.
From Fig.~\ref{fig:3} it can also be seen that the melting behavior is less
cooperative for \Zspa\ than for the Z domain. This conclusion
is supported by our data for the specific heat (not shown); the peak in
the specific heat is more pronounced for the Z domain than
for \Zspa.
 
That the model predicts \Zspa\ to be more compact than the Z domain
is not surprising, given that the number of hydrophobic amino acids is
larger for \Zspa\ (14) than for the Z domain (11). In addition,
\Zspa\ has one more Pro than the wild-type sequence, which does change the
local properties of the chain and could affect the overall size, too.
It should be pointed out that the effect of a Pro on the overall size
may be poorly described by the model because all peptide bonds, also
those preceding a Pro, are held fixed (trans).

The reduced helix content of \Zspa\ shows that this sequence does
not make a perfect three-helix bundle, but does not tell how its
structure differs from a three-helix bundle. It could be that
one of the three helices is missing and that the other two are still there,
but it could also be that the disorder is more uniform along the chain, so
that all three helices are present but partially disordered. The NMR
analysis of \Zspa~\cite{Wahlberg:03} does not exclude any of these two
possibilities.

\begin{figure}[t]
\vspace{0mm}
\begin{center}
\epsfig{figure=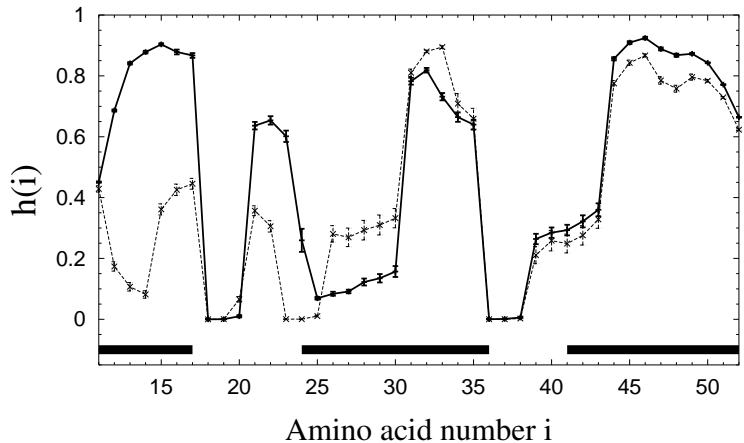,height=6cm}
\end{center}
\vspace{-12pt}
\caption{
Helix content along the chain, $h(i)$, for the 46-amino acid fragments of
\Zspa\ (dashed line) and the Z domain (full line) at $T=0.87\,\Tm$,
where $\Tm$ is the melting temperature for the Z domain.
$h(i)$ denotes the probability that amino acid $i$ is
helical (for the definition of helical, see footnote). Thick
horizontal lines indicate helical parts of the NMR
structure~\protect\cite{Tashiro:97} for the Z domain.}
\label{fig:4}
\end{figure}

Figure~\ref{fig:4} shows how the helix content varies along the chains
in our model. A comparison with experimental data for the Z
domain~\cite{Tashiro:97} shows that the first half of helix II is somewhat
distorted in the model. As a result, it is possible that the model
underestimates the structural
change produced by the mutation Glu25Pro (see Table~\ref{tab:1}),
which should have a helix-breaking effect. Our results for helices I and III
of the Z domain are, by contrast, in good agreement with experimental data.
These two helices respond very differently to the mutations leading
to \Zspa. Our results suggest that helix III, which itself is free from
mutations, remains stable in \Zspa, whereas helix I, which contains seven
mutations (see Table~\ref{tab:1}), turns unstable. Two possible explanations 
why the model predicts helix I to become unstable are that the hydrophobicity
pattern of helix I is less helical in \Zspa, and that one of the mutations,
Phe13Gly, increases the flexibility of this part of the chain.

\begin{figure}
\vspace{0mm}
\begin{center}
\mbox{
\epsfig{figure=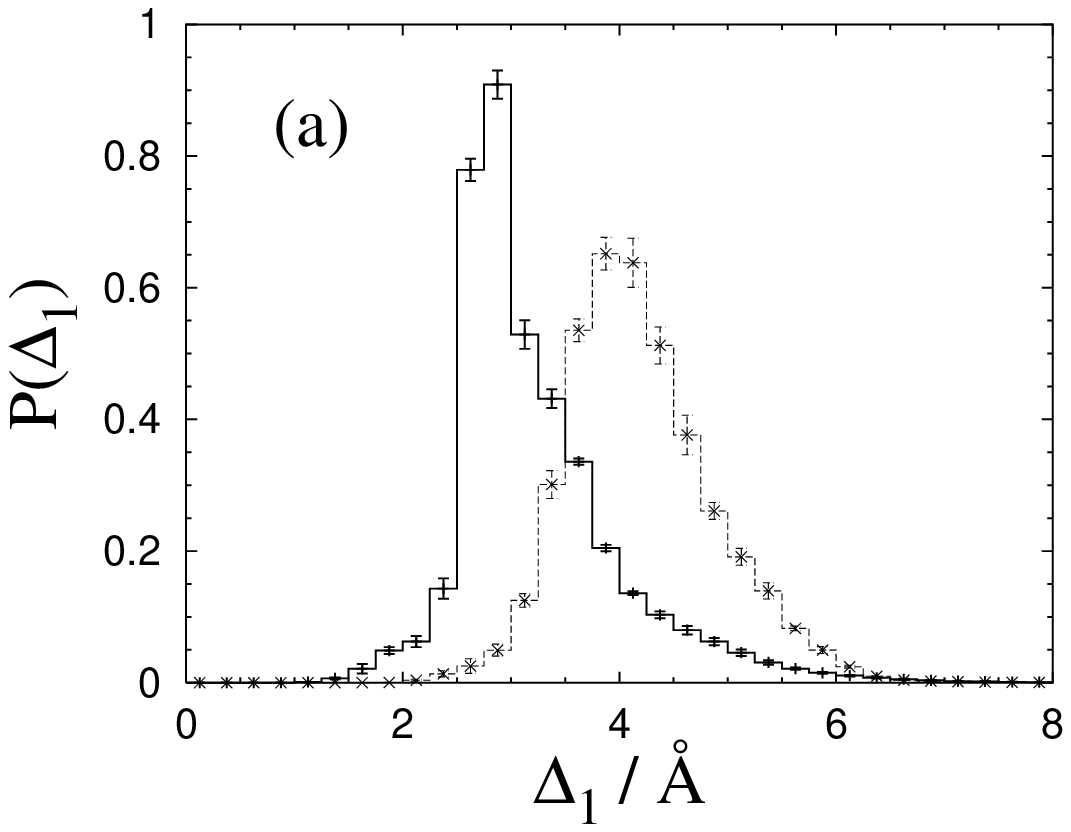,height=5.5cm}
  \hspace{5mm}
\epsfig{figure=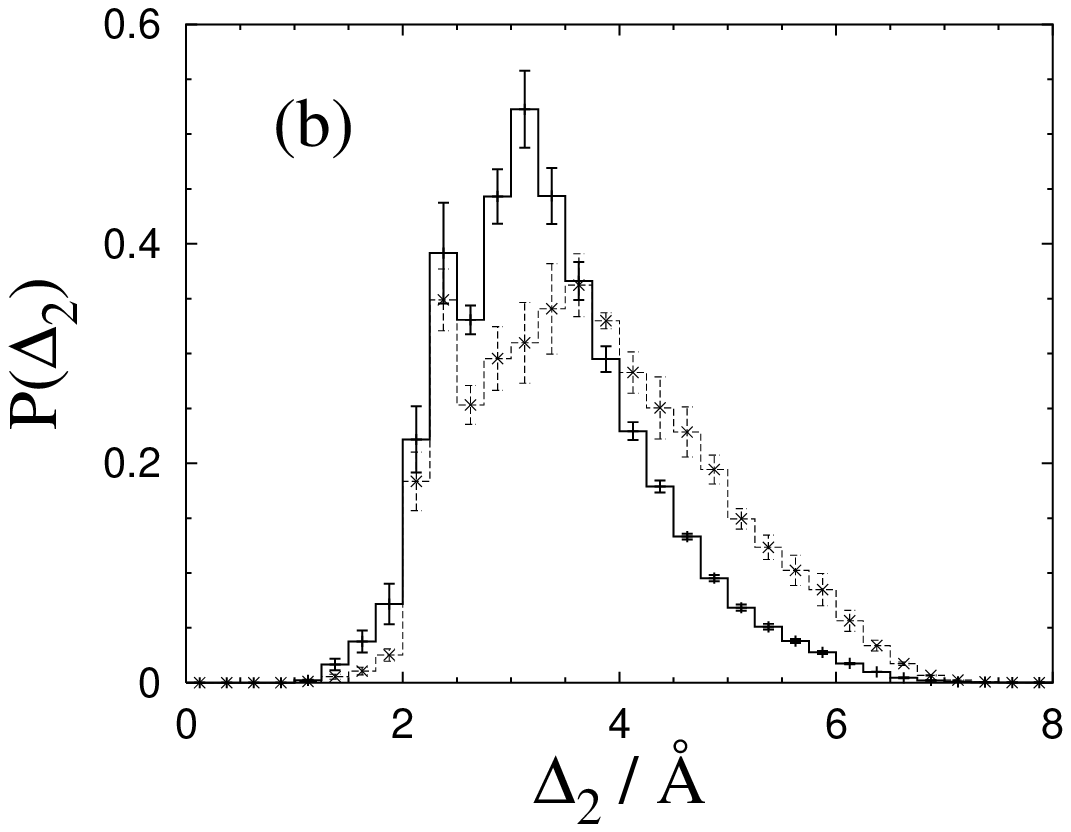,height=5.5cm}
}
\end{center}
\vspace{-12pt}
\caption{
RMSD distributions for the 46-amino acid fragments of \Zspa\ (dashed line)
and the Z domain (full line). (a) The distribution of
$\Delta_1$ (amino acids 9--31). (b) The distribution of
$\Delta_2$ (amino acids 32-54). Both $\Delta_1$ and $\Delta_2$
are backbone RMSDs. The temperature is the same as in Fig.~\ref{fig:4}.}
\label{fig:5}
\end{figure}

To further investigate how the mutations affect different parts
of the chain, we also perform an RMSD-based analysis. For each conformation,
we compute two RMSD values, $\Delta_1$ and $\Delta_2$, for the first
and second halves of the chain, respectively. The two parts of the
chain are separately superimposed on the NMR structure.
Figure~\ref{fig:5} shows the probability distributions of
$\Delta_1$ and $\Delta_2$ both for \Zspa\ and the Z domain.
In line with the results in Fig.~\ref{fig:4}, we find that the
two $\Delta_2$ distributions are similar, although the distribution
for \Zspa\ is slightly wider. The two $\Delta_1$ distributions differ,
by contrast, markedly, the mean being significantly higher
for \Zspa\ than for the wild-type sequence. These results
confirm that, in our model, the disorder of \Zspa\ is not uniformly
distributed along the chain; the main difference between \Zspa\ and
the Z domain lies in the behavior of the first half of the chain.

\section{Conclusion}

Using a model that combines a relatively detailed chain representation
with a simple interaction potential, we have studied the thermodynamic
behaviors of an engineered sequence and its parent. The model is
sequence-based, which makes it possible to compare the two sequences
in a straightforward manner. Despite the simplicity of the potential,
we found that the model is able to capture important effects of
the mutations; the mutated sequence, \Zspa, shows a
reduced helix content and a melting behavior
that is less cooperative, compared with the wild-type
sequence. We also found that chain collapse occurs before helix
formation sets in for \Zspa, and that the main difference between
the two sequences lies in the behavior of the first half of the chain,
which is less stable for \Zspa. To decide whether or not these two
predictions are correct requires further experimental data.

{\large \bf Acknowledgments}

We thank Torleif H\"ard for a helpful discussion. 
This work was in part supported by the Swedish Research Council.

\end{document}